# Sequential toy model for electron transfer in donor-bridge-acceptor systems

Kamil Walczak [1]

Institute of Physics, Adam Mickiewicz University
Umultowska 85, 61-614 Poznań, Poland

Sequential tunneling model is used to study electron transport in molecular rectifiers based on structures of donor-bridge-acceptor type. The device is made of two metallic electrodes connected by a molecule, which contains acceptor and donor subunits, separated by insulating bridge. Both subunits are modeled as quantum dots with discrete energy levels, isolated from each other by potential barrier and weakly coupled to both electrodes through tunnel junctions. Intervalence donor-acceptor tunneling process is treated as a superexchange mechanism or equivalently through the use of Marcus theory. Analytic formula for the current is found in the case of the Aviram-Ratner ansatz of rectification and current-voltage characteristic is obtained at an arbitrary strength of the potential drop over the tunneling region. It is shown that rectification current depends on the position of the acceptor's LUMO and donor's HOMO levels with respect to the Fermi energy of the electrodes before bias is applied, and their shift due to the bias voltage.

## I. Introduction

Rectifiers always have played a key role in the development of molecular electronics (moletronics) as a proposal of the simplest components in future electronic circuits [1]. Such devices are usually composed of donor-bridge-acceptor structures sandwiched between two metallic electrodes. Electrical current is driven under the influence of an external bias. Molecular diodes of that type have been synthesized, and rectifying behavior (i.e. strong asymmetry in the conductance spectrum) has been measured on molecular layers formed by Langmuir-Blodgett (LB) techniques [2-4]. The physical mechanism for the rectification is not clear yet, but few different suggestions has been made [1,5,6]. In general, transport properties of such structures are affected by: the quantum nature of molecular system, electronic properties of the electrodes near the Fermi energy level, and the strength of the molecule-to-electrodes coupling.

The main purpose of this work is to find an analytic formula for the current due to a simple model of rectification. Sequential tunneling model is proposed to describe the conduction in devices based on structures of donor-bridge-acceptor type ($D-\sigma-A$) [7] located in between two macroscopic electrodes ($M_1$ and $M_2$). Here $D$ is good electron-donor subunit with low ionization potential, $A$ is good electron-acceptor with high electron affinity, and $\sigma$ denotes insulating saturated covalent bridge. In this model, molecular subunits ($D$ and $A$) are treated as quantum dots (with discrete energy levels), while connecting bridges are modeled as potential barriers for moving electrons. The existence of tunnel barriers in molecular junctions can be questionable, however such barriers at a metal/molecule interfaces have been reported in the literature [8,9]. Furthermore, without real spatial extent, the voltage is dropped entirely at that barriers.

The unique features of molecular subsystems (donor and acceptor) are ensured by chemical substituent groups bound to a single molecules. Such "intramolecular dopants" can push electrons into or take them from that system, due to their electronegativity. Because of that tendency all the substituents are classified as electron-releasing ($-NH_2$, $-OH$, $-CH_3$, $-OCH_3$, $-CH_2CH_3$) or withdrawing ($=O$, $-NO_2$, $-CN$, $-CHO$, $-CF_3$). So even with no external applied bias, there is a dopant-induced difference in the relative energetic positions of the π-orbitals in the donor and acceptor subunits of the molecular structure. The consequence of such energy-level difference is asymmetry in the electrical current flowing through the device in two opposite polarities of applied bias (diode-like character of the current-voltage dependence).

## II. The Aviram-Ratner ansatz

Mechanism of rectification in analyzed structures is based on the Aviram-Ratner ansatz [1]. The effect of rectification is associated with the fact that the excited zwitterionic state $D^+ - \sigma - A^-$ is relatively accessible from the ground neutral state $D - \sigma - A$, while the opposite zwitterion $D^- - \sigma - A^+$ lies several electronvolts higher in energy and therefore is inaccessible (except at dielectric breakdown) [3]. Forward-biased electron transfer in the Aviram-Ratner mechanism is schematically presented as a two-step process:

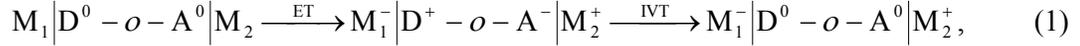

$$M_1 \big| D^0 - o - A^0 \big| M_2 \xrightarrow{ET} M_1^- \big| D^+ - o - A^- \big| M_2^+ \xrightarrow{IVT} M_1^- \big| D^0 - o - A^0 \big| M_2^+, \quad (1)$$

where: ET is elastic tunneling (from the HOMO of the neutral donor molecule onto the metal electrode $M_1$, and simultaneously from the electrode $M_2$ onto the LUMO of the neutral acceptor molecule), while IVT denotes an intervalence transfer (inelastic through-bond tunneling). Similarly, scheme for two-step process of electron transfer under the reverse bias can be written in the form:

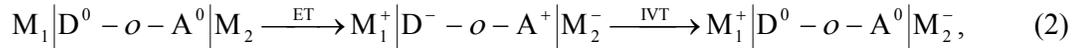

$$M_1 \big| D^0 - o - A^0 \big| M_2 \xrightarrow{ET} M_1^+ \big| D^- - o - A^+ \big| M_2^- \xrightarrow{IVT} M_1^+ \big| D^0 - o - A^0 \big| M_2^-, \quad (2)$$

where: ET is elastic tunneling (from the electrode $M_1$ onto the LUMO of the neutral donor molecule, and simultaneously from the HOMO of the neutral acceptor molecule onto the metal electrode $M_2$). Here we also completely neglect the competing reverse-bias process associated with autoionization, followed by charge migration to the metallic electrodes [1].

## III. Current formula

Here we assume that the electronic states of particular segments of the molecule (donor and acceptor parts) are perfectly isolated from each other (no hybridization) due to insulating spacer (bridge). Since only HOMO and LUMO levels are the most important in the transport description, we focus our attention on the mentioned energy levels only. General formula for the current flowing through the device can be written as follows (see the Appendix):

$$I(V) = I_0 \int_{-\infty}^{+\infty} d\varepsilon' \int_{-\infty}^{+\infty} d\varepsilon f(\varepsilon, \mu_S) \zeta(\varepsilon - \varepsilon_L^A)[1 - f(\varepsilon', \mu_{Dr})] \zeta(\varepsilon' - \varepsilon_H^D)$$

$$- I_0 \int_{-\infty}^{+\infty} d\varepsilon' \int_{-\infty}^{+\infty} d\varepsilon f(\varepsilon, \mu_{Dr}) \zeta(\varepsilon - \varepsilon_L^D)[1 - f(\varepsilon', \mu_S)] \zeta(\varepsilon' - \varepsilon_H^A). \quad (3)$$

In the above double-integration procedure (3): $f(x)$ is the equilibrium Fermi distribution function and $\zeta(x)$ is the so-called broadening function of molecular energy level, which is assumed to be of Lorentzian shape:

$$\zeta(\varepsilon - \varepsilon^{A/D}) = \frac{1}{2\pi} \frac{\Gamma_{S/Dr}}{(\varepsilon - \varepsilon^{A/D})^2 + \frac{1}{4}\Gamma_{S/Dr}^2}. \quad (4)$$

The strength of the coupling between the molecule and the metallic electrodes (due to the thiol linkages) is described by widths parameters $\Gamma_S$ and $\Gamma_{Dr}$, which are assumed to be energy and voltage independent [10-12]. Both electrodes are treated as semi-infinite reservoirs of free electrons at thermal equilibrium, which are characterized through the electrochemical potentials: $\mu_{S/Dr} = \varepsilon_F \pm eV/2$.

Furthermore, the magnitude of the current flowing through the device is proportional to a prefactor $I_0$, which takes into account the transfer rate between acceptor and donor subunits (multiplied by an elementary electronic charge $e$):

$$I_0 = \frac{e}{\tau_s} \exp[-\beta R_{DA}]. \quad (5)$$

Here: $\beta$ is a structure-dependent attenuation factor, $R_{DA}$ is the distance between donor and acceptor species (the length of a bridge) and $\tau_s$ is donor-acceptor tunneling time. Such time $\tau_s$ is approximately equal to experimentally-determined time involved in bridge vibrations and generally can be temperature-dependent. When donor and acceptor subsystems are weakly coupled with the help of $\sigma$-bonded bridge, the electron transfer takes place directly from donor to acceptor via tunneling process (also referred to as superexchange mechanism) and exhibits exponential decay as a function of distance [13,14].

Instead of superexchange theory, it is also possible to describe intervalence donor-acceptor tunneling process with the help of Marcus theory [15]. In this case of a classical description of the transfer rate, the expression for prefactor $I_0$ is given by [16]:

$$I_0 = e \frac{k_B T}{h} \exp\left[-\frac{\Delta G^*}{k_B T}\right], \quad (6)$$

where: $k_B$ is Boltzmann constant, $T$ is absolute temperature and $\Delta G^*$ is the activation free energy associated with donor-acceptor electron transfer reaction. However, both approaches indicate that the reduction of the current is caused mainly by some exponential factor.

**IV. An example**

Now we proceed to apply presented formalism to the case of proposed model of molecular diode (shown in Fig.1). Let us assume that such device is composed of donor $C_6H_4(NH_2)_2$ and acceptor $C_6H_4(CHO)_2$ subunits, separated by insulating $-(CH_2)_3-$ bridge. Each end of the molecule is attached to the surface of the metallic electrode with the help of $-S-$ terminal atoms in the chemisorption process [17]. In general, the electronic structure of molecular systems can be calculated by various techniques, including: empirical, semiempirical, *ab initio* or density functional approaches, which are based on different degrees of approximation. The planar geometry for the bare molecule was determined by energy optimization using the standard parametrization of the ZINDO/1 package implemented in Hyperchem [18]. To obtain energy levels of isolated segments (donor and acceptor), we have performed calculations using Extended Hückel Theory (EHT) method once the geometry of the molecule is known. This choice is dictated because of the absence of adjustable parameters in EHT scheme and for the sake of simplicity (basis set includes all of the orbitals of valence electrons, completely neglecting electron-electron interactions). Hence, energy levels for both subsystems are (given in eV): $\varepsilon_H^D = -11.290$, $\varepsilon_L^D = -8.346$, $\varepsilon_H^A = -12.647$, $\varepsilon_L^A = -10.549$. Since the voltage is dropped entirely at the tunnel barriers, the energies of the molecular states depend on the applied bias through the relations: $\varepsilon^A = \varepsilon - eV/6$, $\varepsilon^D = \varepsilon + eV/6$. For saturated hydrocarbon chains $\beta = 0.9$ Å$^{-1}$ [19-21], the length of the bridge $-(CH_2)_3-$ is $R_{DA} = 5$ Å and molecular vibration period for that bridge is of order of $\tau_s \sim 10^{-12}$ s [22]. Fermi energy of the electrodes is usually located somewhere in between the HOMO-LUMO gap [10,23], and in this work it is arbitrarily chosen to be equal to $\varepsilon_F = -10.816$ eV. To simulate bad contacts with the electrodes (as a typical situation in experiments) we take the parameters responsible for that coupling as: $\Gamma_S = \Gamma_{Dr} = 0.02$ eV.

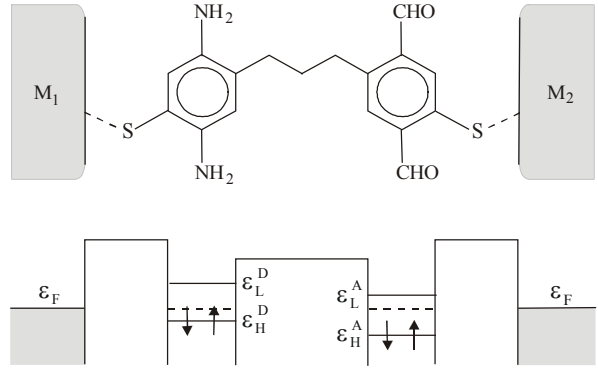

Fig.1 Model of $D-\sigma-A$ molecular rectifier:
schematic representation of the device
and its energy level diagram.

Figure 2 shows the current-voltage (I-V) characteristic for molecular diode proposed in this section (depicted in Fig.1) at temperature T = 290 K. Such type of molecular junction rectifies current in the forward direction (positively biased) relatively to that the reverse direction (negatively biased). Strong asymmetry of the I-V dependence is due to asymmetry in the structure of the molecule itself. Substituting all the parameters indicated in the previous paragraph, an estimation of the current in the usual units is: $I_0 = 1.8$ nA. In order to obtain the same result from Marcus theory we should assume that: $\Delta G^* = 0.1575$ eV. Character of the I-V curve and the evaluation of the peak value of the current are similar to some experimental results [2-4].

## V. Concluding remarks

Presented model qualitatively reproduces the rectifier behavior of complex molecular devices. The magnitude of the predicted current is comparable with experimental data in the conditions that we match all the parameters. So rectification is attributed to asymmetry of a molecule with an acceptor-bridge-donor structure ($A-\sigma-D$), where mechanism of diode-like behavior results from the Aviram-Ratner ansatz. However, this toy model seems to be useful as a simple theoretical tool in designing molecular devices of the donor-bridge-acceptor type, which demonstrate rectifying features. One of a crucial problems stems from the task of finding Fermi energy of the electrodes (while experimental control of this level is realizable through the choice of metal, which play the role of the electrode).

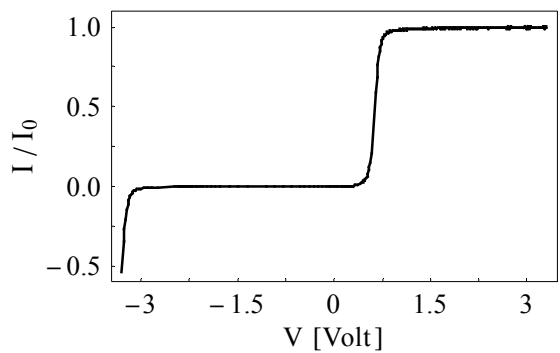

Fig.2 Rectification curve in the I-V characteristic
for analyzed model of molecular diode.

Moreover, the tunneling of an electron through the acceptor-bridge-donor junction is treated as a sequential process, in which the molecular species are successively charged and discharged. In fact, after the initial tunneling step an electron can have enough time to redistribute charge along the molecule (i.e. to polarize this molecule). Such phenomenon can have an influence on the shifts of molecular energy levels, and transfer-induced dipole momentum can affect transport characteristics. However, here we do not include this effect into our formalism, completely ignoring such phenomena like Coulomb blockade. In the frames of this work we also neglected all the effects of electron correlations as well as inelastic scattering.

It should be also noted that several experiments on symmetric molecules have shown rectification [8,9,24,25], while others have found very little or no rectification [26,27]. Such effect is justified by asymmetric electrode coupling, which can result in an asymmetric potential drop along the length of the molecule. If the potential profile is asymmetric, molecular energy level through which an electron is propagated can line up differently in positive and negative bias, resulting in rectification [27-30]. Rectifying features of molecular systems can also be explained as a consequence of conformational changes due to: thermal effects [31], inelastic tunneling [32] or application of electric field [33].

**Acknowledgments**

The author is very grateful to Ben Janesko for illuminating suggestions.

**Appendix**

Electric current in the forward (source-to-drain) direction can be expressed as a product of probabilities of electron source-to-acceptor transfer and hole drain-to-donor transfer, and tunneling acceptor-to-donor transfer rate (multiplied by an elementary electronic charge e ):

$$I_f = eP_e(S \to A)P_h(Dr \to D)k_T(A \to D), \qquad (A.1)$$

where $k_T$ can be modeled within superexchange or Marcus theory, respectively. Similarly electric current in the backward (drain-to-source) direction is given by the analogous product relation:

$$I_b = eP_e(Dr \to D)P_h(S \to A)k_T(D \to A). \qquad (A.2)$$

The general expression for the current is given finally by the difference between the forward- and the backward- flowing currents: $I = I_f - I_b$. The problem we are facing now is associated with definitions of all the probabilities of electron (hole) transfer from the electrode to the particular species of the molecule (donor and acceptor, respectively). Here we assume that such quantities are directly proportional to the probabilities of an electron existing in the electrode at the molecular energy levels involved in transport process (as modified by the coupling with the electrodes):

$$P_e(S \to A) = \int_{-\infty}^{+\infty} d\varepsilon f(\varepsilon, \mu_S)\zeta(\varepsilon - \varepsilon_L^A), \qquad (A.3)$$

$$P_h(Dr \to D) = \int_{-\infty}^{+\infty} d\varepsilon [1 - f(\varepsilon, \mu_{Dr})]\zeta(\varepsilon - \varepsilon_H^D), \qquad (A.4)$$

$$P_e(Dr \to D) = \int_{-\infty}^{+\infty} d\varepsilon f(\varepsilon, \mu_{Dr})\zeta(\varepsilon - \varepsilon_L^D), \qquad (A.5)$$

$$P_h(S \to A) = \int_{-\infty}^{+\infty} d\varepsilon [1 - f(\varepsilon, \mu_S)]\zeta(\varepsilon - \varepsilon_H^A). \qquad (A.6)$$

In the above relations: $\varepsilon$ is the injection energy of the tunneling electron, $f(x)$ denotes the Fermi distribution function and $\zeta(x)$ is the broadening function (in which all the information regarding to the coupling with the electrodes is included).

# References

[1] E-mail address: walczak@amu.edu.pl
[1]  A. Aviram, M. A. Ratner, Chem. Phys. Lett. **29**, 277 (1974).
[2]  N. J. Geddes, J. R. Sambles, D. J. Jarvis, W. G. Parker, D. J. Sandman, J. Appl. Phys. **71**, 756 (1992); R. M. Metzger, Mat. Sci. Eng. C **3**, 277 (1995); S. Scheib, M. P. Cava, J. W. Baldwin, R. M. Metzger, Thin Solid Films **327-329**, 100 (1998); R. M. Metzger, Acc. Chem. Res. **32**, 950 (1999).
[3]  R. M. Metzger *et al.*, J. Am. Chem. Soc. **119**, 10455 (1997).
[4]  D. Vuillaume, B. Chen, R. M. Metzger, Langmuir **15**, 4011 (1999); B. Chen, R. M. Metzger, J. Phys. Chem. B **103**, 4447 (1999); R. M. Metzger, J. Mater. Chem. **9**, 2027 (1999), J. Mater. Chem. **10**, 55 (2000), Synth. Met. **109**, 23 (2000); R. M. Metzger, T. Xu, I. R. Peterson, Angew. Chem. Int. Ed. Engl. **40**, 1749 (2001), J. Phys. Chem. **105**, 7280 (2001).
[5]  J. C. Ellenbogen, J. C. Love, Proc. IEEE **88**, 386 (2000).
[6]  C. Krzeminski, C. Delerue, G. Allan, D. Vuillaume, R. M. Metzger, Phys. Rev. B **64**, 085405 (2001).
[7]  C. Majumder, H. Mizuseki, Y. Kawazoe, J. Phys. Chem. A **105**, 9454 (2001).
[8]  C. Zhou, M. R. Deshpande, M. A. Reed, L. Jones II, J. M. Tour, Appl. Phys. Lett. **71**, 611 (1997).
[9]  C. Kergueris, J.-P. Bourgoin, S. Palacin, D. Esteve, C. Urbina, M. Magoga, C. Joachim, Phys. Rev. B **59**, 12505 (1999).
[10] L. E. Hall, J. R. Reimers, N. S. Hush, K. Silverbrook, J. Chem. Phys. **112**, 1510 (2000).
[11] S. T. Pantelides, M. Di Ventra, N. D. Lang, Physica B **296**, 72 (2001).
[12] T. Kostyrko, B. Bułka, Phys. Rev. B **67**, 205331 (2003).

[13] W. B. Davis, W. A. Svec, M. A. Ratner, M. R. Wasielewski, Nature **396**, 60 (1998).
[14] G. Pourtois, D. Beljonne, J. Cornil, M. A. Ratner, J. L. Bredas,
     J. Am. Chem. Soc. **124**, 4436 (2002).
[15] R. A. Marcus, Chem. Phys. Lett. **133**, 471 (1987); *ibid*. **146**, 13 (1988).
[16] K. V. Mikkelsen, M. A. Ratner, Chem. Rev. **81**, 113 (1987).
[17] A. Ulman, Chem. Rev. **96**, 1533 (1996), and references cited therein.
[18] HYPERCHEM 5.1 Pro for Windows (Hypercube Inc., 1997).
[19] H. Oevering M. N. Paddon-Row, M. Heppener, A. M. Oliver, E. Cotsaris,
     J. W. Verhoeven, N. S. Hush, J. Am. Chem. Soc. **109**, 3258 (1987).
[20] M. D. Johnson, J. R. Miller, N. S. Green, G. L. Closs, J. Phys. Chem. **93**, 1173 (1989).
[21] S. B. Sachs, S. P. Dudek, R. P. Hsung, L. R. Sita, J. F. Smalley, M. D. Newton,
     S. W. Feldberg, C. E. D. Chidsey, J. Am. Chem. Soc. **119**, 10563 (1997).
[22] M. Olson, Y. Mao, T. Windus, M. Kemp, M. Ratner, N. Léon, V. Mujica,
     J. Phys. Chem. B **102**, 941 (1998).
[23] W. Tian, S. Datta, S. Hong, R. Reifenberger, J. I. Henderson P. Kubiak, J. Chem. Phys.
     **109**, 2874 (1998); S. N. Yaliraki, A. E. Roitberg, C. Gonzalez, V. Mujica,
     M. A. Ratner, J. Chem. Phys. **111**, 6997 (1999); M. Di Ventra, S. T. Pantelides,
     N. D. Lang, Phys. Rev. Lett. **84**, 979 (2000); P. S. Damle, A. W. Ghosh, S. Datta,
     Phys. Rev. B **64**, R201403 (2001).
[24] M. Pomerantz, A. Aviram, R. A. McCorkle, L. Li, A. G. Schrott,
     Science **255**, 1115 (1992).
[25] A. Dhirani, P.-H. Lin, P. Guyot-Sionnest, R. W. Zehner, L. R. Sita,
     J. Chem. Phys. **106**, 5249 (1997).
[26] M. A. Reed, C. Zhou, C. J. Muller, T. P. Burgin, J. M. Tour, Science **278**, 252 (1997).
[27] S. Datta, W. Tian, S. Hong, R. Reifenberger, J. I. Henderson, C. P. Kubiak,
     Phys. Rev. Lett. **79**, 2530 (1997).
[28] J. Taylor, M. Brandbyge, K. Stokbro, Phys. Rev. Lett. **89**, 138301 (2002).
[29] V. Mujica, M. A. Ratner, A. Nitzan, Chem. Phys. **281**, 141 (2002).
[30] P. E. Kornilovitch, A. M. Bratkovsky, R. S. Williams, Phys. Rev. B **66**, 165436 (2002).
[31] Z. J. Donhouser, B. A. Manthooth, K. F. Kelly, L. A. Bumm, J. D. Monnel,
     J. J. Stapleton, D. W. Price, A. M. Rowlett, D. L. Allara, J. D. Tour, P. S. Weiss,
     Science **292**, 2303 (2001).
[32] J. Gaudioso, L. J. Lauhon, W. Ho, Phys. Rev. Lett. **85**, 1918 (2000).
[33] A. Troisi, M. A. Ratner, J. Am. Chem. Soc. **124**, 14528 (2002).